# Second harmonic generation in surface periodically-poled lithium niobate waveguides: on the role of multiphoton absorption

M. Cherchi,[1] S. Stivala[1,2], A. Pasquazi,[2] A. C. Busacca,[1] S. Riva Sanseverino,[1,3] A. C. Cino,[1,3] L. Colace,[2] and G. Assanto [2]

[1] *DIEET, Università di Palermo, Viale delle Scienze, 90128 Palermo, ITALY*

[2] **NooEL**-*Nonlinear Optics and OptoElectronics Lab, University Roma Tre, Via della Vasca Navale 84, 00146 Roma, ITALY*

[3] *CRES - Centro per la Ricerca Elettronica in Sicilia, Via Regione Siciliana 49, 90046 Monreale (PA), ITALY*

**Abstract**

Second harmonic generation is investigated in lithium niobate channels realized by proton exchange and quasi-phase-matched by surface periodic-poling. The reduction in conversion efficiency at high powers is interpreted in terms of multi-photon absorption via two-color terms, yielding an estimate of the dominating three-photon process.

32.80.Wr, 42.65.-k , 42.65.Ky , 42.65.Wi, 42.70.Mp, 77.84.Dy, 78.20.Bh

cherchi@unipa.it, FAX +39 091 6406200

# Introduction

Guided-wave configurations allow to enhance intensity dependent phenomena in optical materials [1-2]. Lithium niobate (LN) is one of the most used dielectrics for optoelectronics and harmonic generation. The latter often relies on quadratic response and ferroelectric properties of LN, enabling techniques for periodic poling and quasi-phase matched (QPM) parametric interactions [3]. Recently, improved technologies and pulsed sources have fostered a better understanding of the interplay and the combined effects of second and higher-order optical nonlinearities, including the role of cascading phenomena through up- and down-conversion [4-11]. In particular, with explicit reference to surface periodic poling in lithium niobate waveguides [12], we recently highlighted the combined effects of cubic and $\chi^{(2)}$-cascaded interactions in low-yield second harmonic generation (SHG) [13]. For high intensities above, however, even the role of nonlinear absorption (NLA) can become manifest [14-18]. In this Paper we report some recent results on second harmonic generation in LN channel waveguides realized by proton exchange (PE) and surface periodic poling. As we observed a reduction in conversion efficiency (defined as the ratio between the input power in the fundamental mode and the output power in the second harmonic mode) at high excitation powers, we undertook the study of the detrimental effects of NLA, identifying the dominant processes and estimating the size of three-photon absorption in LN.

## 2. Surface Poled Waveguides and Setup

We conducted our experiments in surface periodically poled (SPP) [12] lithium niobate with proton exchanged [19] channel waveguides, as sketched in Fig. 1. We employed an optical parametric amplifier/oscillator pumped by an amplified and frequency doubled pulse train and by a single pulse, both coming from a picosecond Nd:YAG laser. The output is tunable in the range 1064-2200 nm with less than 2 cm$^{-1}$ linewidth, 25 ps pulse width and 10 Hz repetition rate. After being "cleaned up" by a spatial filter, the laser beam is focused by a 750 mm focal length lens and its waist is about 305 μm.

The fundamental frequency (FF) laser beam was end-fire coupled into the channel waveguide with a spot of 3.6μm; the outputs at fundamental and second harmonic modes were imaged with a Vidicon tube and a CCD camera, respectively; powers were monitored with Ge and Si photodiodes with the aid of boxcar averager. The temperature was stabilized around room value during all measurements.

By launching peak powers of tens of kW launched in 1cm-long waveguides; we were able to resolve SHG spectral features at excitations high enough to induce nonlinear losses while avoiding the detrimental effects of photorefractive damage. Neither we expected free-carrier generation (unlike in Ref. 14). It can be also easily checked that (unlike in Ref. 20) neither the low energy of single pump pulses (below $0.5\,\mu J$), nor the low average launched power (below 10 W/cm$^2$) could induce significant thermooptic effect.

QPM was implemented with $16.8\,\mu m$ periodicity (designed for phase matching around 1550 nm wavelength) in $500\,\mu m$-thick Z-cut congruent LN wafers via SPP [12], and the X-propagating waveguides from 1 to $7\,\mu m$ in width were realized by proton exchange in benzoic acid with 3% lithium benzoate through a photo-lithographically defined mask [13, 19-22]. The resulting $\alpha$-phase [23] waveguides (surface extraordinary index increase $\approx 1.2\times 10^{-2}$) were about $2.6\,\mu m$ deep and supported a single (weakly guided) TM mode around 1.55μm [24] with an effective area of $53\,\mu m^2$ and a single even TM mode at 770 nm with an effective area of $23\,\mu m^2$ (both values calculated from the experimentally acquired modal distributions). Morphology and profile of the ferroelectric domains after (over)poling via SPP and PE were revealed by selective chemical etching, electron microscopy and surface analysis; the un-inverted domains were shallow (less than 1μm deep) and only partially overlapped with the guided eigenmodes, thereby reducing the strength of the QPM SHG process [13,26].

## 3. Experimental Results

As previously reported in Ref. 13, SHG in 4 μm wide waveguides proved to be the most efficient (about 8% conversion efficiency when launching 10 KW), despite the limited depth of the periodically (un)poled domains (consistently with the morphologic analysis, our numerical fits corresponded to less than 450 nm deep domains) due to an unoptimized SPP process. Owing to light propagation over LN regions without QPM, Kerr effect together with quadratic cascading due to phase mismatch induced appreciable SHG resonance shift in wavelength at FF peak powers of the order of 10 kW [13]. When we increased the pump peak power above 15 kW, however, the conversion efficiency started to decrease at a rate higher than predicted by the coupled nonlinear differential equation model including quadratic and cubic nonlinearities (Fig. 2.a), although the shift could be correctly modeled (Fig. 2.b) (the fitting parameters of these curves are the same as in the improved model that will be presented in details afterwards). While some degree of sub-linear growth of the conversion efficiency is due to spectral features inherent to the use of temporal pulses and self-phase modulation via an effective Kerr effect of cubic and quadratic origin, the excess decrease observed in the experiments indicates contributions from the imaginary parts of the nonlinear cubic –or even quintic- susceptibility. Temporal walk-off and group velocity dispersion, in fact, can be entirely neglected in our samples due to the duration of the pulses [13]. Moreover, the linear wavelength shift of Fig. 2.b makes unlikely any role of free carrier generation, as the latter would be associated to a quadratic shift [14]. Finally, the linear trend of the shift suggested ruling out photorefractive damage [25], as we verified by increasing the launched power and checking stability over time and repeatability of the of the output efficiency and of the output modal profiles. Nonlinear absorption was thereby considered as the most likely cause of the reduced efficiency at high input powers.

## 4. Nonlinear absorption and SHG

Nonlinear absorption during parametric generation can stem from various multiphoton transitions driven by the imaginary parts of odd-order susceptibilities [17,27]. Writing the electric field at a given waveguide section as

$$E_{tot} = \frac{1}{2}(A_\omega \exp(i\omega t) + A_{2\omega} \exp(2i\omega t) + \text{c.c.}), \quad (1)$$

with $\omega$ being the fundamental frequency and $A_\omega$ ($A_{2\omega}$) the fundamental (second harmonic) complex amplitude. Self- and cross-phase terms from the third order polarization and for both the harmonics are (we neglect the tensorial nature of the nonlinearity as we deal only with linearly polarized fields along the optic axis z)

$$\begin{cases} P_\omega^{(3)} = \frac{3}{8}\varepsilon_0 \chi^{(3)}\left[\left(|A_\omega|^2 A_\omega + 2|A_{2\omega}|^2 A_\omega\right)\exp(i\omega t) + \text{c.c.}\right] \\ P_{2\omega}^{(3)} = \frac{3}{8}\varepsilon_0 \chi^{(3)}\left[\left(|A_{2\omega}|^2 A_{2\omega} + 2|A_\omega|^2 A_{2\omega}\right)\exp(2i\omega t) + \text{c.c.}\right] \end{cases}, \quad (2)$$

where $\varepsilon_0$ is the vacuum permittivity. The terms above correspond to the transitions summarized in Fig. 3, their imaginary parts accounting for two-photon absorption (2PA), which grows linearly with intensity. Terms $A_\omega$ and $A_{2\omega}$ correspond respectively to annihilated photons of frequency $\omega$ and $2\omega$, while their complex conjugated $A_\omega^*$ and $A_{2\omega}^*$ correspond to created $\omega$ and $2\omega$ photons. For the $n^{\text{th}}$ order polarization, all these terms are multiplied by $(1/2)^n \varepsilon_0 \chi^{(n)}$ and by a factor $n!/(n_1! \cdots n_k!)$ (taking into account that $k$ amplitude factors may be repeated $n_1,..., n_k$ times). Thereby, for the fifth order polarization terms at frequencies $\omega$ and $2\omega$:

$$\begin{cases} P_\omega^{(5)} = \dfrac{5}{16}\varepsilon_0 \chi^{(5)} \left[\left(|A_\omega|^4 A_\omega + A_{2\omega}^2 (A_\omega^*)^3 + 6|A_\omega|^2 |A_{2\omega}|^2 A_\omega + 3|A_{2\omega}|^4 A_\omega\right)\exp(i\omega t) + \text{c.c.}\right] \\ P_{2\omega}^{(5)} = \dfrac{5}{16}\varepsilon_0 \chi^{(5)} \left[\left(|A_{2\omega}|^4 A_{2\omega} + 3|A_\omega|^4 A_{2\omega} + \tfrac{1}{2} A_\omega^4 A_{2\omega}^* + 6|A_\omega|^2 |A_{2\omega}|^2 A_{2\omega}\right)\exp(2i\omega t) + \text{c.c.}\right] \end{cases}, (3)$$

stemming from the transitions in Fig. 4 and featuring 2-, 3-, or 4-photon absorption processes. These terms grow with the square of the intensity or the product of two intensities. 2PA and 4PA mixing terms are highly phase mismatched, i.e., they can be neglected in the guided wave configuration. For the same reason we also neglected the contributions to the NLA coming from the (imaginary part of) second-, third-, fourth-, and fifth-order susceptibilities through various frequency mixing processes. Phase matching considerations rule out also multi step absorption processes, like green induced infrared absorption [28], that could come, in principle, from green light generated by sum frequency generation between the pump and the second harmonic.

For the monochromatic case we can write the imaginary part of the nonlinear refractive index in the form $k = k_1 + k_2 I + k_3 I^2$ ($I$ being the optical intensity). From the monochromatic version of (2) and (3), it is straightforward to show perturbatively that [29] $k_2 = 3\operatorname{Im}(\chi^{(3)})/(4\varepsilon_0 c n^2)$ and $k_3 = 5\operatorname{Im}(\chi^{(5)})/(4\varepsilon_0 c^2 n^3)$, where $\varepsilon_0$ is the vacuum permettivity and $n$ is the real part of the linear refractive index. In this way it is possible to define the 2PA coefficient $\beta = 4\pi k_2/\lambda$ and the 3PA coefficient $\gamma = 4\pi k_3/\lambda$.

The diagrams in Figs. 3 and 4 enable to select those terms which are able to contribute to NLA depending, to a first approximation, on the position of the upper energy level compared to the ultraviolet absorption edge of LN. Owing to phonon contributions, even those processes laying in the tail just below the edge may play a significant role [15,16,30]. Since the LN energy gap is in the range (3.8÷3.9) eV [30,31], the only transitions with energy above the edge are the three fifth-order 3PA processes leading to the absorption of at least two SH photons. Concerning the cubic terms, even though the $2\omega + 2\omega$ absorption energy is too low, we can regard it as belonging to the tail just below the edge. In fact not only third order cross sections are much higher than fifth order cross sections, but an indirect gap transition of about 3.28 eV was reported in LN [30]. So we cannot *a priori* exclude a significant contribution arising from this term. In the absence of estimates for either the 2PA coefficient $\beta$ at 770 nm or the 3PA $\gamma$ at any wavelengths in LN, we tried to best fit our experimental results to yield upper bounds to 2PA and 3PA coefficients, comparing the results against experimental values in literature. In particular, $\beta$ values reported at wavelengths just above the edge [15,17,31] are of the order of $10^{-12}$ m/W; since phonon-assisted processes should be at least four orders of magnitude smaller [16], our fit is sensible if the found $\beta$ is well below $10^{-12}$ m/W. As far as 3PA, measured $\gamma$ in various solids range between $10^{-29}$ m$^3$/W$^2$ and $10^{-25}$ m$^3$/W$^2$ and are of the order of $10^{-26}$ m$^3$/W$^2$ in photorefractive crystals [17].

Ref. [32] reports qualitative evidence of 3PA in LN at 780 nm, without a quantitative evaluation of $\gamma$.

## 5. Model and simulations

First-order quasi phase matching with inverted domains of limited depth can be effectively modelled [13, 29, 33] by the following two coupled differential equations for the FF field $w(x,t)$ and the second harmonic field $v(x,t)$ (normalized such that their square moduli are the modal powers in W)

$$\frac{\partial w}{\partial x} = SH_w + PM_w + LA_w + NLA_w$$
$$\frac{\partial v}{\partial x} = SH_v + PM_v + LA_v + NLA_v \quad (4)$$

having defined the second order terms

$$SH_w = -w^* v [\chi_1 \exp(-i\Delta\beta_1 x) + i\chi_2 \exp(-i\Delta\beta_2 x)]$$
$$SH_v = w^2 [\chi_1 \exp(i\Delta\beta_1 x) - i\chi_2 \exp(i\Delta\beta_2 x)] \quad (5)$$

the phase modulation terms proportional to the real part of the third order nonlinear refractive index $n_2$ (measured in $m^2/W$)

$$PM_w = -i n_2 \frac{2\pi}{\lambda_w} (f_{ww} |w|^2 + 2 f_{wv} |v|^2) w$$
$$PM_v = -i n_2 \frac{4\pi}{\lambda_w} (f_{vv} |v|^2 + 2 f_{vw} |w|^2) v \quad (6)$$

the linear absorption terms $LA_u = -\alpha_u/2\, u$ ($u = w,v$, where the linear absorption coefficients $\alpha_u$ account for both material and propagation losses), and, consistently with the previous paragraph, starting from Eq. (2) and (3), we have calculated the NLA terms, proportional to the imaginary parts of the third and fifth order nonlinear refractive indexes $k_2$ and $k_3$ respectively,

$$NLA_w = -k_3 \frac{6\pi}{\lambda_w} f_{wvv}^2 |v|^4 w$$
$$NLA_v = -k_2 \frac{4\pi}{\lambda_w} f_{vv} |v|^2 v - k_3 \frac{4\pi}{\lambda_w}(f_{vvv}^2 |v|^4 + 6 f_{vvw}^2 |v|^2 |w|^2) v \quad (7)$$

Defining the modal wave vectors $\beta_i \equiv 2\pi n_i / \lambda_i$ and the $\Lambda$-period grating wave vector $k_G \equiv 2\pi/\Lambda$, in Eq. (5) the unpoled region is governed by the mismatch $\Delta\beta_2 \equiv \beta_v - 2\beta_w$ and the poled regions by the mismatch $\Delta\beta_1 \equiv \Delta\beta_2 - k_G$, while the nonlinear coefficients $\chi_i$ (i=1,2) are defined as

$$\chi_i \equiv d_{eff_i} g_i \sqrt{\frac{8\pi^2 f_{SHG}}{c\varepsilon_0 \lambda_w^2 n_w^2 n_v}} \quad (8)$$

$n_i$ ($i = w, v$) being the modal effective index of the two modes, and having defined the effective nonlinear coefficients $d_{eff_1} = 2 d_{33}/\pi$ and $d_{eff_2} = d_{33}$, and the overlap integral

$$f_{SHG} \equiv \frac{\left[\int_{-\infty}^{+\infty}\int_{-\infty}^{+\infty} e_v^*(y,z) e_w^2(y,z) dy dz\right]^2}{\left[\int_{-\infty}^{+\infty}\int_{-\infty}^{+\infty} |e_w(y,z)|^2 dy dz\right]^2 \int_{-\infty}^{+\infty}\int_{-\infty}^{+\infty} |e_v(y,z)|^2 dy dz}, \quad (9)$$

where $e_i(x,y)$ ($i = w, v$) are the transverse distributions of the waveguide modes. Defining $z_0$ as the coordinate of the boundary between the two regions, the weighting parameters $g_i$ are defined as

$$g_1 \equiv \frac{\int_{-\infty}^{+\infty}\int_{z_0}^{+\infty} e_v^*(y,z)e_w^2(y,z)dydz}{\int_{-\infty}^{+\infty}\int_{-\infty}^{+\infty} e_v^*(y,z)e_w^2(y,z)dydz} \quad \text{and} \quad g_2 = 1 - g_1. \quad (10)$$

The self- and cross-phase modulation overlap integrals in Eq. (6) and Eq. (7) are defined as

$$f_{jk} \equiv \frac{\int_{-\infty}^{+\infty}\int_{-\infty}^{+\infty}|e_j(y,z)|^2|e_k(y,z)|^2 dydz}{\int_{-\infty}^{+\infty}\int_{-\infty}^{+\infty}|e_j(y,z)|^2 dydz \int_{-\infty}^{+\infty}\int_{-\infty}^{+\infty}|e_k(y,z)|^2 dydz}$$

$$f_{jkl} \equiv \sqrt{\frac{\int_{-\infty}^{+\infty}\int_{-\infty}^{+\infty}|e_j(y,z)|^2|e_k(y,z)|^2|e_l(y,z)|^2 dydz}{\int_{-\infty}^{+\infty}\int_{-\infty}^{+\infty}|e_j(y,z)|^2 dydz \int_{-\infty}^{+\infty}\int_{-\infty}^{+\infty}|e_k(y,z)|^2 dydz \int_{-\infty}^{+\infty}\int_{-\infty}^{+\infty}|e_l(y,z)|^2 dydz}} \quad . \quad (11)$$

Consistently with the experimental evidence, we neglected the contributions from the real part of the fifth order refractive index. Their presence, in fact, would be in contrast with the linear trend of the measured wavelength shift shown in Fig. 2.b.

Starting from low power throughput data at both wavelength, and taking in to account the the Fresnel relations together with the overlap integral between the laser mode and the guided modes, we evaluated propagation losses about 0.2 cm$^{-1}$ for both modes and laser-to-waveguide input coupling of 73%. We used a standard WKB technique to reconstruct the index profile of the waveguides from the experimental data, and a commercial mode solver [34] enabled us to estimate the effective index dispersion of both FF and SH. The overlap integrals (11) have been calculated from the experimentally acquired modal distributions, resulting in SHG effective area $1/f_{SHG} = 76.68\,\mu m^2$ , self phase modulation effective areas $1/f_{ww} = 52.99\,\mu m^2$, $1/f_{vv} = 23.11\,\mu m^2$, and $1/f_{vvv} = 19.96\,\mu m^2$, and cross phase modulation effective areas $1/f_{wv} = 44.08\,\mu m^2$ and $1/f_{wvv} = 28.22\,\mu m^2$.

Finally, we assumed a gaussian intensity profile for the input pulses, and, consistently to what found in our previous work [13], a quadratic nonlinear coefficient $d_{33} = 18$ pm/V with a domain depth of 430 nm, and a nonlinear refractive index $n_2 = 5 \times 10^{-20}\,m^2/W$ [35]. Actually we could have chosen any other couple of quadratic and cubic nonlinear coefficients that fitted the low power data (e.g. $d_{33} = 16.5$ pm/V and $n_2 = 10 \times 10^{-20}\,m^2/W$) since this choice cannot affect the fitting analysis of NLA, as it is clear from the form of the differential equations (4).

We numerically integrated the coupled equations and then we integrated the solutions on the whole range of instantaneous values, trying to fit our data with NLA coefficients. In a first attempt we assumed no fifth-order contributions, in order to check whether the 2PA had a leading role. The results, shown in Fig. 5, correspond to a coefficient $\beta = 1.1 \times 10^{-12}\,m/W$ at 770 nm, that physically means $\beta = (1.1 \pm 0.1) \times 10^{-12}\,m/W$, when considering a 10% relative error owing to the systematic relative error (about 5%) in evaluating the laser-to-waveguide input coupling. However, this $\beta$ appears far too large at this wavelength, as it is comparable to those reported at 532 nm, i.e. well above the LN absorption edge. In fact, the only available 2PA estimate close to 770 nm [15] is an upper limit of about $6 \times 10^{-14}\,m/W$ (as the experimental procedure could not discriminate between 2PA and 3PA). In any case, even assuming this limit as the correct value, 2PA could give a maximum 6% contribution to NLA, well within the experimental error.

The results of simulations assuming only 3PA are displayed in Fig. 6, where we set $\gamma = 4.5 \times 10^{-28} \, \text{m}^3/\text{W}^2$ at 770 nm, that again physically means, taking into account the systematic errors, $\gamma = (4.5 \pm 0.7) \times 10^{-28} \, \text{m}^3/\text{W}^2$. This best fit value is well within the range discussed in the previous paragraph. From Eq. (7), NLA is dominated by the bichromatic term with the product of two intensities (corresponding to the last process in Fig. 4), nearly two orders of magnitude larger than the other ones. Notice that the quintic terms just below the UV edge, even in the hypothesis of 3PA two orders of magnitude smaller than found above the edge, would contribute at most at 3%, i.e. within the experimental error.

## 6. Conclusions

During ps SHG in SPP-LN waveguides we observed that the efficiency peak decreased for input intensities $> 30 \, \text{GW/cm}^2$. Having experimentally and theoretically excluded any contribution of free carrier generation, photorefractive damage, thermo-optic effect and multi step absorption processes, we were led to explain this trend in terms of NLA from third and fifth-order self-and cross-phase processes. Based on numerical simulations and fitting, a significant contribution of 2PA resulted in contrast with previously reported values of the 2PA coefficient , so that we could estimate the 3PA coefficient in LN α-phase PE SPP waveguides. While this estimate is rather indirect and we cannot directly extrapolate this value to pure crystals, this appears to be the first experimental evidence of bichromatic 3PA in lithium niobate.

**Acknowledgements**

This work was funded by the Italian Ministry for Scientific Research through PRIN 2005098337 and, in part, through project APQ RS-19 (CIPE 17/2002, BCNanolab). We thank Prof. He Ping Li for useful discussions and D. Maxein and Prof. K. Buse for making their 2PA measurements in LN available to us.

**Figure Captions**

1. Fig. 1 Sketch of the nonlinear waveguide. The LN crystal is poled everywhere but retains a shallow periodic pattern that doesn't reach the whole PE channel depth.

2. Fig. 2 (a) Measured (open circles) and predicted (solid line) maximum SHG efficiency versus launched FF peak power. (b) Measured (open circles) and predicted (solid line) wavelength shift. Nonlinear absorption is not included in the model.

3. Fig. 3 Diagrams of cubic self- and cross-phase transitions.

4. Fig. 4 Diagrams of quintic self- and cross-phase and frequency mixing transitions.

5. Fig. 5 Maximum efficiency versus peak input power: experimental values (open circles) are numerically interpolated (solid curve) assuming 2PA only.

6. Fig. 6 Maximum efficiency versus peak input power: experimental values (open circles) are numerically interpolated (solid curve) assuming 3PA only.

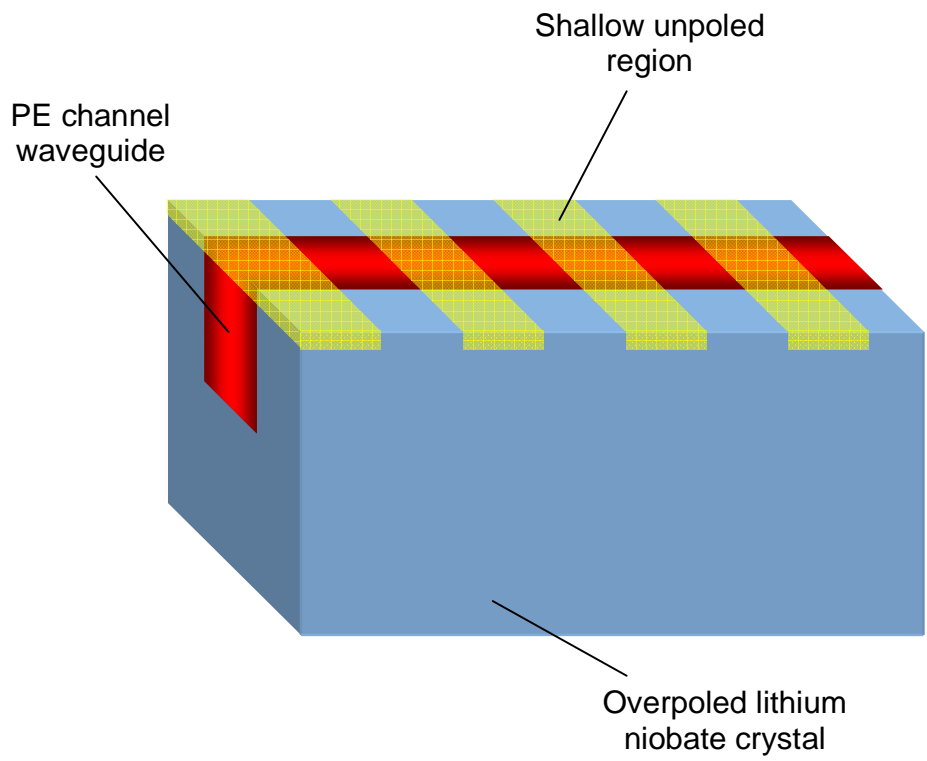

Fig. 1 Sketch of the nonlinear waveguide. The LN crystal is poled everywhere but retains a shallow periodic pattern that doesn't reach the whole PE channel depth.

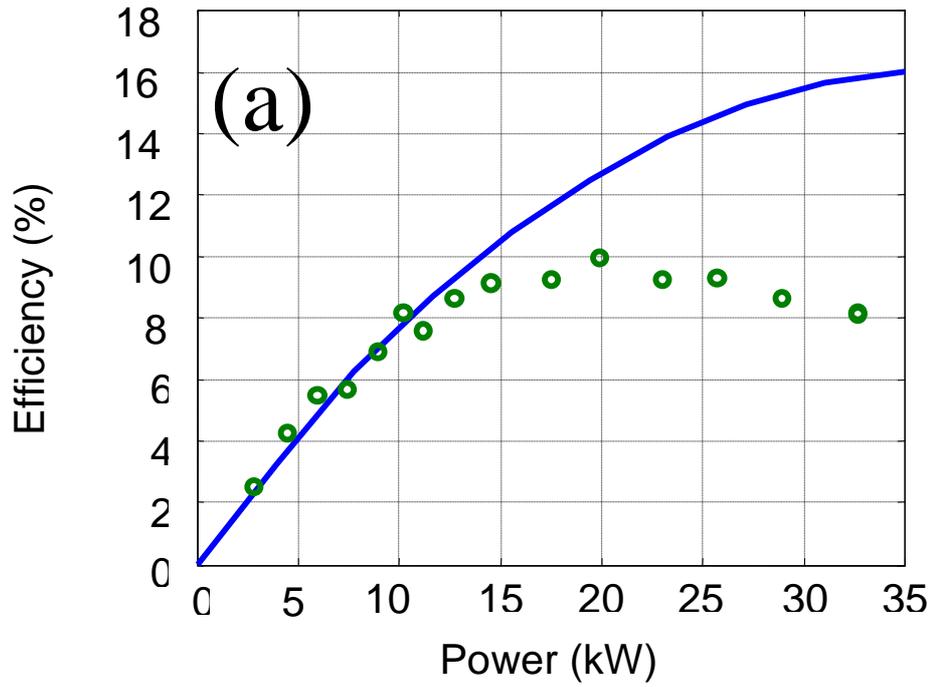

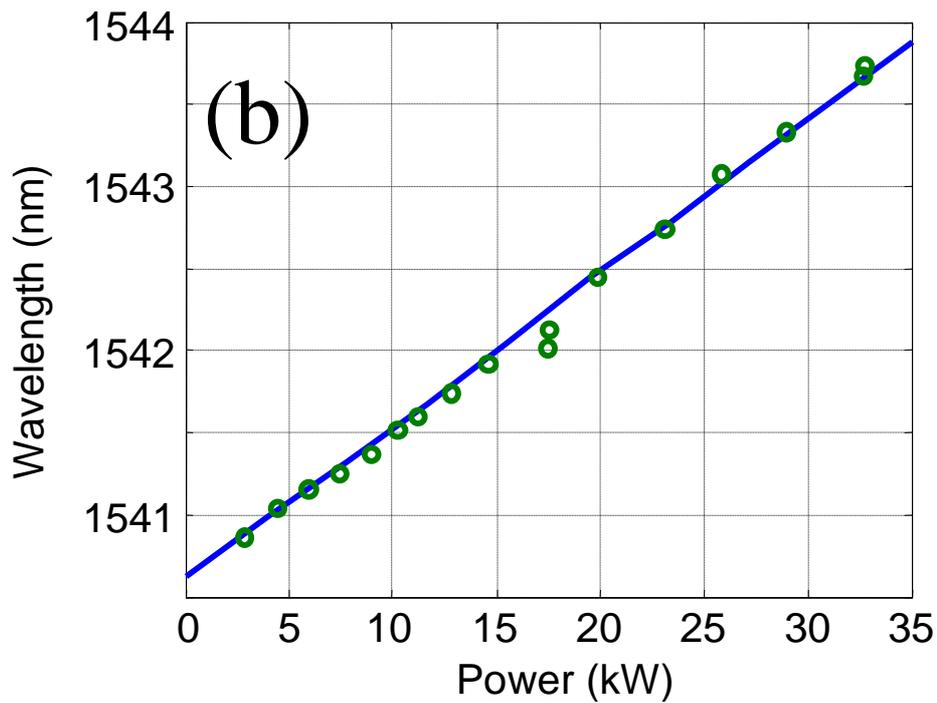

Fig. 2 (a) Measured (open circles) and predicted (solid line) maximum SHG efficiency versus launched FF peak power. (b) Measured (open circles) and predicted (solid line) wavelength shift. Nonlinear absorption is not included in the model.

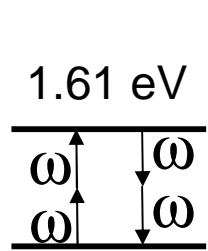
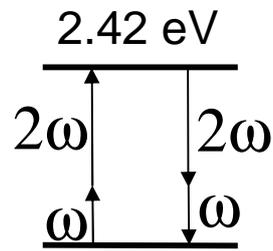
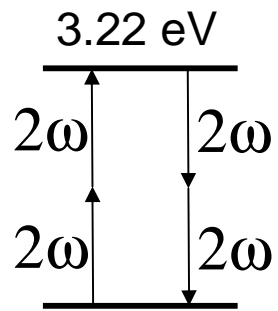
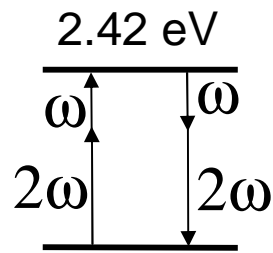

Fig. 3 Diagrams of cubic self- and cross-phase transitions.

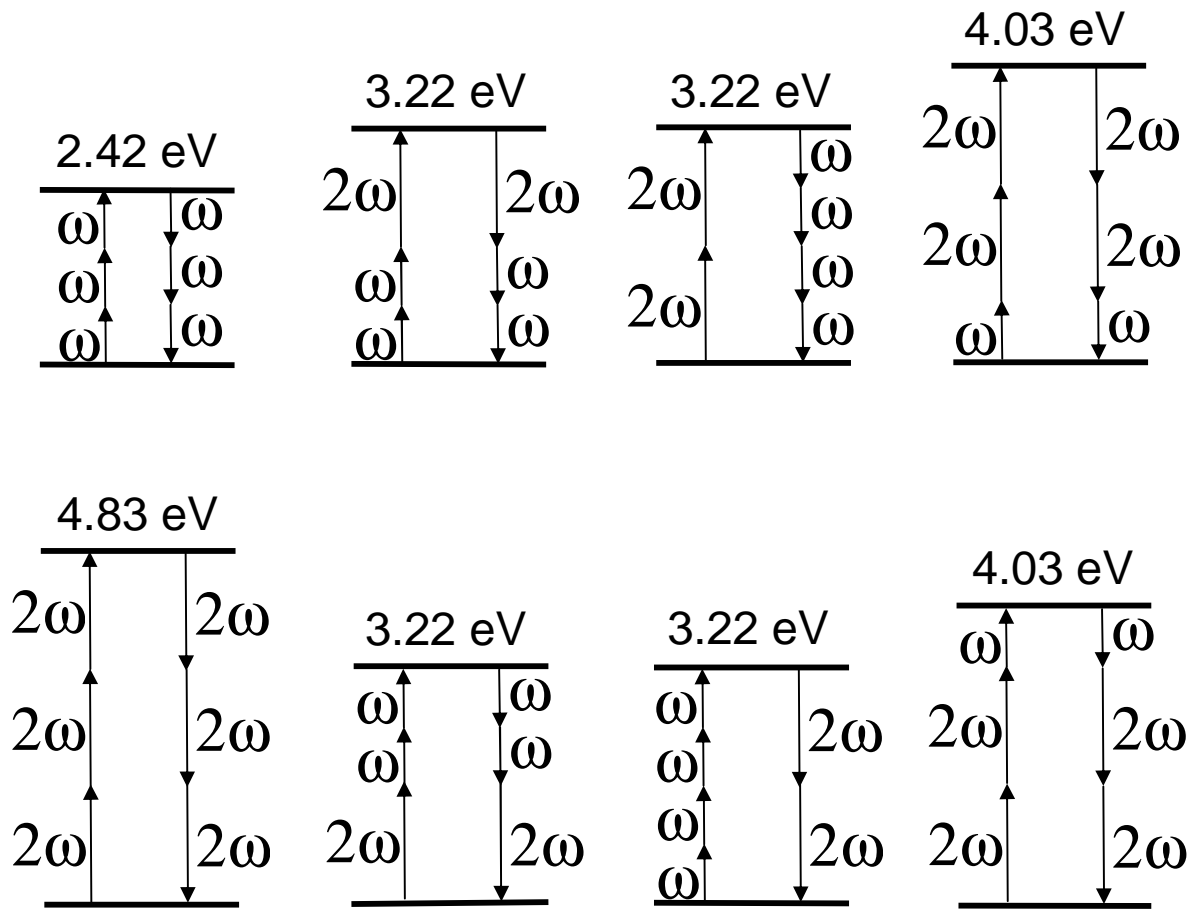

Fig. 4 Diagrams of quintic self- and cross-phase and frequency mixing transitions.

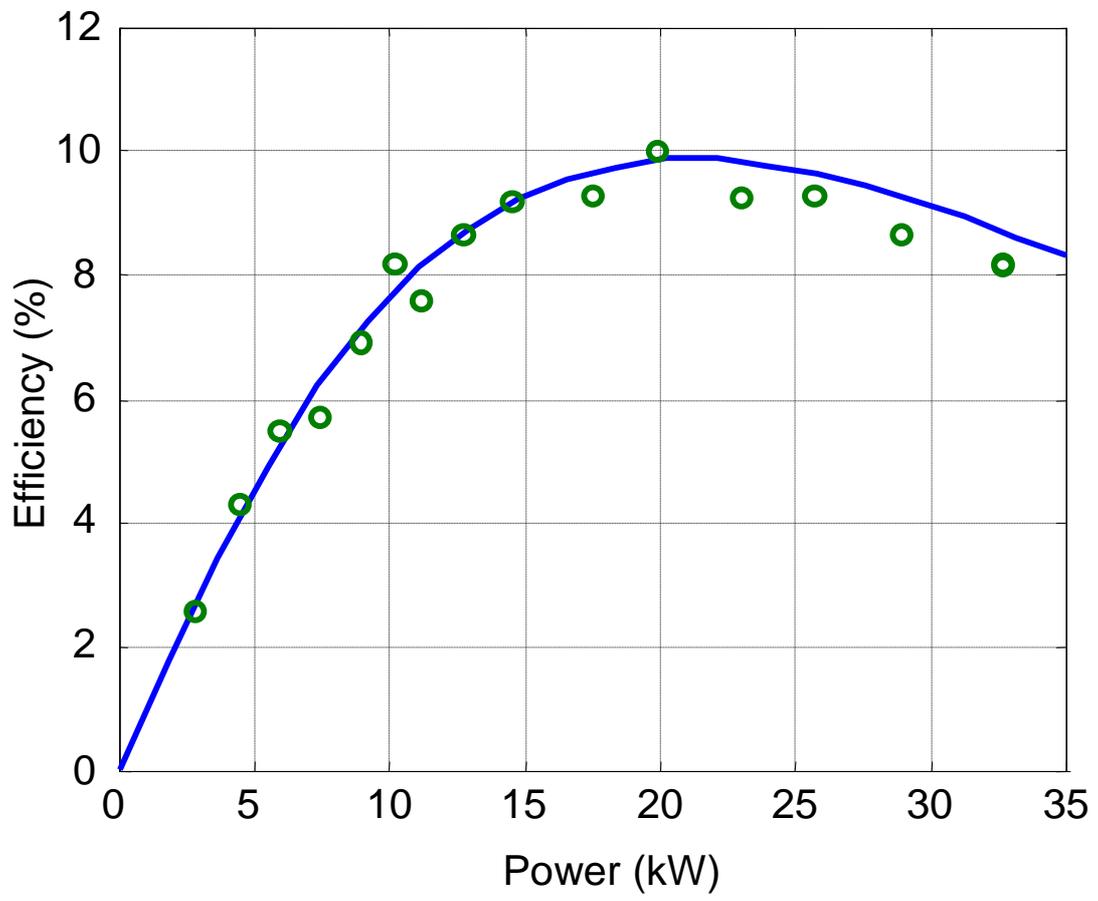

Fig. 5 Maximum efficiency versus peak input power: experimental values (open circles) are numerically interpolated (solid curve) assuming 2PA only.

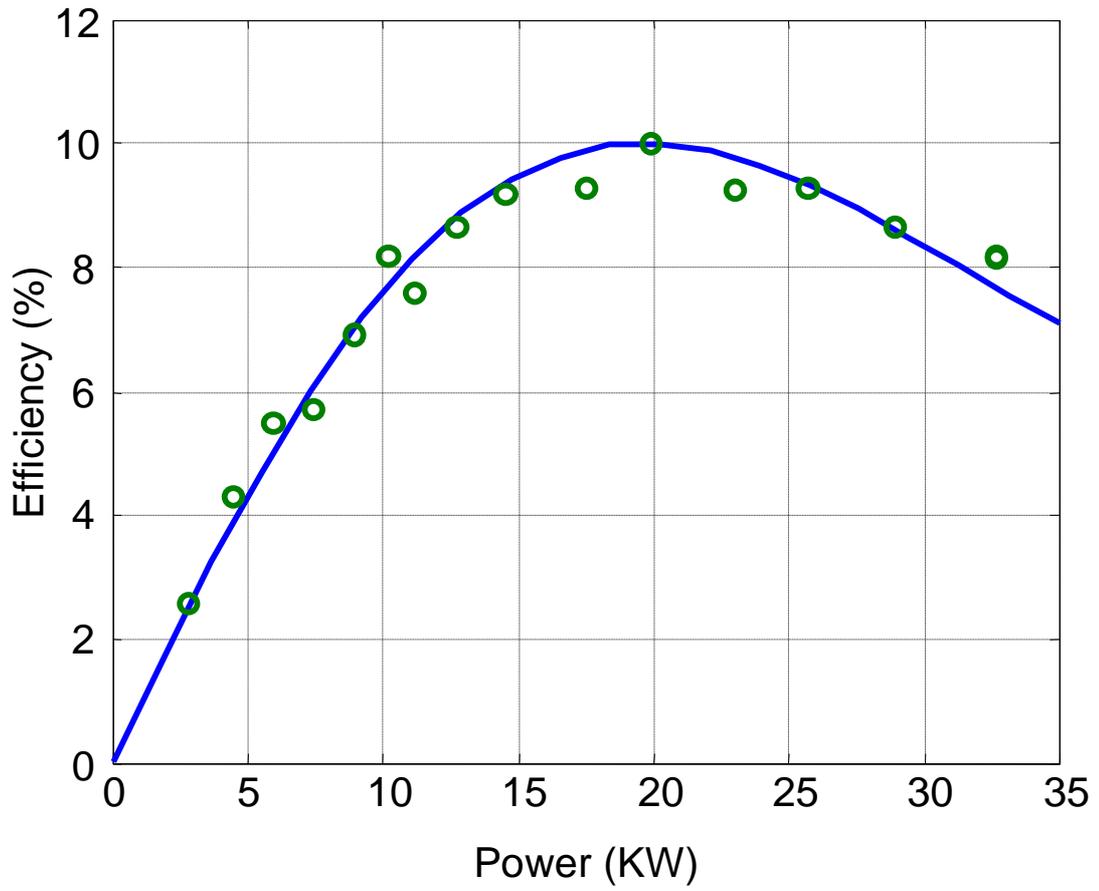

Fig. 6 Maximum efficiency versus peak input power: experimental values (open circles) are numerically interpolated (solid curve) assuming 3PA only.